\documentclass[conference]{IEEEtran}
\IEEEoverridecommandlockouts
\usepackage{cite}
\usepackage{amsmath,amssymb,amsfonts}
\usepackage{algorithmic}
\usepackage{graphicx}
\usepackage{textcomp}
\usepackage{xcolor}
\usepackage{booktabs}
\usepackage{mdnotation}

\def\BibTeX{{\rm B\kern-.05em{\sc i\kern-.025em b}\kern-.08em
    T\kern-.1667em\lower.7ex\hbox{E}\kern-.125emX}}
\begin{document}

\title{Candidate Set Sampling for Evaluating Top-$N$ Recommendation\\
\thanks{This work is partially supported by the National Science Foundation under Grant IIS 17-51278.}
}

\author{\IEEEauthorblockN{Ngozi Ihemelandu}
\IEEEauthorblockA{\textit{Department of Computer Science} \\
\textit{Boise State University}\\
Boise, USA \\
ngoziihemelandu@u.boisestate.edu}
\and
\IEEEauthorblockN{Michael D. Ekstrand}
\IEEEauthorblockA{\textit{Department of Computer Science} \\
\textit{Boise State University}\\
Boise, USA \\
ekstrand@acm.org \\
\IEEEauthorblockA{\textit{(Now at Drexel University)}}}
}

\maketitle

\begin{abstract}
The strategy for selecting candidate sets --- the set of items that the recommendation system is expected to rank for each user --- is an important decision in carrying out an offline top-$N$ recommender system evaluation.
The set of candidates is composed of the union of the user's test items and an arbitrary number of non-relevant items that we refer to as decoys. 
Previous studies have aimed to understand the effect of different candidate set sizes and selection strategies on evaluation.
In this paper, we extend this knowledge by studying the specific interaction of candidate set selection strategies with popularity bias, and use simulation to assess whether sampled candidate sets result in metric estimates that are less biased with respect to the true metric values under complete data that is typically unavailable in ordinary experiments.
\end{abstract}


\section{Introduction}
\label{intro}
Recommender systems are evaluated online using techniques such as A/B testing, as well as offline using historic data and evaluation performance metrics for tasks such as rating prediction and top-$N$ recommendation. In the majority of published research, algorithms are compared offline \cite{cremonesi2021progress}. 

Offline evaluations require multiple experimental design decisions; one of these is the strategy for selecting candidate sets (also known as target item sampling \cite{canamares2020offline}).
The candidate set ($C_{\User}$) is the set of items that the recommendation system is expected to rank for each user in a top-$N$ experiment. The default practice is to select all items except items that the user has interacted with ($\Items \setminus \Items[\User]^{\Ltrain}$), where~$\Items$ is the set of all items and~$\Items[\User]^{\Ltrain}$ is the set of items for which a training rating by user $u$ is available. In this paper, we refer to all items in $C_{\User}$ that are not in $\Items[\User]^{\Ltest}$ as \emph{decoys} ($\Items[\User]^{\Ldecoy}$) \cite{ekstrand2017sturgeon}. $\Items[\User]^{\Ltest}$ is the set of items for which a test rating by user $u$ is available. Therefore, $C_{\User} = \Items[\User]^{\Ltest} \cup  \Items[\User]^{\Ldecoy}$.
Candidate sets that are composed of the union of test items and an arbitrary number of randomly-sampled non-relevant items -- ($\Items[\User]^{\Ltest} \cup \Items[\User]^{\Ldecoy}$) -- are sometimes seen in evaluation reports \cite{koren2008factorization, zhang2018coupledcf, cheng2021dual, ebesu2018collaborative}. 

There have been studies that aimed to understand the effect of different candidate set sizes and selection strategies \cite{canamares2020offline, ekstrand2017sturgeon, bellogin2011precision}. We extend this knowledge by studying specific interaction of candidate set selection strategies with popularity bias. 

Popular items are more likely to be seen by users, rated by users, and are recommended even more frequently than its usefulness as a signal in recommendation warrants. This impacts logged data used for training and evaluating recommender systems such that a small fraction of popular items accounts for most of the user-item interactions \cite{abdollahpouri2020multi}. Hence, recommender system data typically follow long-tailed or power-law distributions.  Popularity bias occurs when evaluation metrics that use test data for evaluation favor recommender systems that tend to recommend popular items, and may penalize recommenders that produce less popular but more novel recommendations that may also be relevant. 

Ekstrand and Mahant \cite{ekstrand2017sturgeon} analytically demonstrated that randomly sampled decoys exacerbates popularity bias in evaluation. Evaluation metrics generally assess the recommender’s ability to separate test items from decoy items in the candidate set.  Test items are (approximately) drawn from the popularity-weighted distribution under most train-test splitting schemes. When decoy items are drawn uniformly, they are drawn from a different distribution that places much less weight on popularity, so popularity is a more useful signal for detecting which distribution a particular candidate item came from. However, they did not observe evidence of the exacerbation of popularity bias effect in their empirical experiments. 

In this paper, we explore this further by empirically investigating the combined effect of the dispersion of interactions (e.g. ratings) across items, and the different candidate set selection strategies on popularity bias. We analyze the effects to determine which factors worsens popularity bias, and which mitigates it. We specifically look into three strategies:
\begin{itemize}
    \item uniformly-sampled decoys
    \item popularity-weighted sampled decoys
    \item full candidate set ($C_{\User} = \Items[\User]^{\Ltest} \cup \Items[\User]^{\Ldecoy} = \Items \setminus \Items[\User]^{\Ltrain}$)
\end{itemize}

Even if we can reduce popularity bias exacerbation in the sampling strategy, that is not enough to demonstrate whether sampled candidate sets are useful.
Krichene et al. \cite{krichene2022sampled} argued that they are not, as metrics computed from sampled candidate sets (which they call "sampled metrics") often do not agree with the metrics computed with full candidate sets.
In terms of statistical estimators, it means that the sampled metric is not an unbiased estimator of the metric computed with full candidate sets.
However, it is not clear that this is the correct estimand: recommender systems evaluation is subtantially affected by the fact that evaluation data is missing-not-at-random \cite{steck2010training}.
Tian and Ekstrand \cite{tian2020estimating} used simulation to generate both complete and observed preference data, allowing the metrics over observed data --- as would be computed in a real experiment --- to be compared with the equivalent metric values if complete evaluation data were available.
We apply this approach to examine whether computing metrics over observed data using sampled or full candidate sets results in a better estimator of the effectiveness metric values complete data would yield.
This is, we argue, a better assessment of the usefulness of sampled metrics for predicting real-world performance.

To understand the usefulness of candidate set sampling we address the following questions:
\begin{enumerate}
    \item [$\textbf{RQ1}$] Does uniform sampling of decoys exacerbate popularity bias in evaluation compared to full candidate sets?
    \item [$\textbf{RQ2}$] Does popularity-weighted sampling of decoys mitigate popularity bias exacerbation?
    \item [$\textbf{RQ3}$] Does uniform or popularity-weighted sampling of the decoys improve the estimation of effectiveness  of recommender system in terms of the bias of the computed metric with respect to the value computed over complete relevance data?
\end{enumerate}

We show that popularity bias is exacerbated when the popularity concentration of the dataset is high, the recommender algorithm has a high propensity to recommend popular items, and the decoys are sampled uniformly. We find that popularity-weighting of the decoys may help mitigate this exacerbation.

We find that with the full candidate set, observed data significantly underestimates the effectiveness measure for all algorithms. The estimation improves with the uniformly sampled, and popularity-weighted candidate sets indicating that candidate set sampling is indeed useful for evaluation.

\section{Background and Related Work}

It is common practice to evaluate recommender systems for top-$N$ recommendation task, where the recommender system is required to suggest a few items to users that they would find appealing and relevant \cite{cremonesi2010performance}.   Their performance of this task is measured with metrics such as nDCG, recall, and hit rate.

The recommender system ranks items from a candidate set. The candidate set is made up of the union of two disjoint sets:  the test items, which the user has rated in the test data, and the decoy items, which are items the user has not rated ($C_{\User} = \Items[\User]^{\Ltest} \cup \Items[\User]^{\Ldecoy}$). 
Historically, the most typical configuration has been to use all items that the user did not rate in the training data as the candidate set ($C_{\User} = \Items \setminus \Items[\User]^{\Ltrain}$).

Instead of considering all items as candidates for ranking, Koren \cite{koren2008factorization} used candidate sets for each user that were made up of $1000$ randomly selected decoys (or non-relevant) test items and one rated test item. In other words, for each test item $i$, rated  by user $u$, an additional $1000$ random items were selected and the ratings by $u$ for $i$ predicted as well as for the other $1000$ unrated items. They wanted to find the relative place of the test items within the total order of test items sorted by predicted ratings for a specific user. They provided no justification for using $1000$ decoys.

Bellog{\'\i}n et al. \cite{bellogin2017statistical} considered candidate set sampling as an alternative design in their study of the implication of applying information retrieval methodologies to recommender system evaluation. They considered different designs of candidate set sampling of which Koren's\cite{koren2008factorization} approach was one.  They analyzed the effect of the number of unrated items as a configuration parameter in candidate set sampling on sparsity, and metric values. They also considered the effect of popularity distribution skewness on popularity bias. However, in this study, we specifically consider the combined effect of the popularity concentration of the dataset, and the candidate set selection strategy on popularity bias.

As we discussed in section \ref{intro}, Krichene et al.\cite{krichene2022sampled} argued that metrics computed from sampled candidate sets often do not agree with the metrics computed with full candidate sets. They proposed metric corrections to attenuate this effect, taking full rankings as the reference

Ca{\~n}amares and Castells\cite{canamares2020target} confirmed some of Krichene et al.'s\cite{krichene2022sampled} findings  that systems are ranked differently when metrics are computed from sampled candidate sets vs full candidate sets. However, they showed that using a full candidate set may not necessarily be the preferable option, and a candidate set that is too small may indeed weaken the reliability and informativeness of the comparison between systems. They also sought to give guidance or understanding of what an appropriate decoy size would be and which size may result in more informative evaluation. 

Like Ca{\~n}amares and Castells\cite{canamares2020target}, Ekstrand and Mahant\cite{ekstrand2017sturgeon} sought to determine how to select an appropriate decoy size. They showed analytically that uniformly-sampled random decoys exacerbates popularity bias, but did not observe empirical evidence of this effect in their experiments. This motivates our work of empirically studying the effect that the sampling of decoys has on popularity bias in evaluation.

Candidate set sampling is a more common practice in sequential recommendation models \cite{latifi2022sequential}.

The contributions of this research are: understanding and mitigating the amplification of popularity bias caused by uniform candidate set sampling, and assessing whether it improves the accuracy of estimating system effectiveness as it would be measured with complete relevance data. 

\section{Candidate Sampling and Popularity Bias}
In this section, we seek to address $\textbf{RQ1}$ and $\textbf{RQ2}$. 

\subsection{Data}
To get the evaluation data needed for the study, we performed an offline evaluation of top-$N$ recommendations while varying the candidate set selection strategy. 
We used three publicly-available data sets (two from MovieLens \cite{harper2015movielens} --- \textit{ML-100K} and \textit{ML-25M} ---  and \textit{Yahoo! R3} from Yahoo! Webscope \cite{marlin2009collaborative}), summarized in Table \ref{tbl:datasets}. The \textit{Yahoo! R3} dataset consists of ratings for music entered by users in the Yahoo! LaunchCast streaming service. It includes: missing not at random ($\operatorname{MNAR}$) training ratings, and missing at random ($\operatorname{MAR}$) test ratings. 

For each dataset:
\begin{itemize}
    \item we split the set of test users into $5$ subsets.
    \item for each subset of test users:
    \begin{itemize}
        \item we select $20\%$ of a test user's interactions for testing, and use the rest for training.
    \end{itemize}
\end{itemize}

For the \textit{Yahoo! R3}, we only split the $\operatorname{MNAR}$ dataset(training ratings) into test and train datasets. We then use this test dataset to compute the regular biased metric, and the $\operatorname{MAR}$ test ratings to compute unbiased metric value.

\begin{table}
    \caption{Summary of data sets.}
    \label{tbl:datasets}
    \centering
    \resizebox{\columnwidth}{!}{
    \begin{tabular}{cccccc}
        Dataset&\# Ratings&\# Users&\# Items & Density & Gini-Index\\
        \toprule
        \textsf{ML-100K} & 100,000 & 1,000 & 1,682 & 6.3\% & 0.6290\\
        \textsf{ML-25M} & 25,000,000 & 162,000 & 59,047 & 0.26\% & 0.8922\\
        \textsf{Yahoo!R3 - Train} & 300,000 & 15,400 & 1,000 & 1.9\% & 0.5595 \\
        \textsf{Yahoo!R3 - Test} & 54,000 & 5,400 &1,000 & 1\% & 0.0798 
    \end{tabular}
    }
\end{table}

\begin{figure}
    \centering
    \includegraphics[width=4cm]{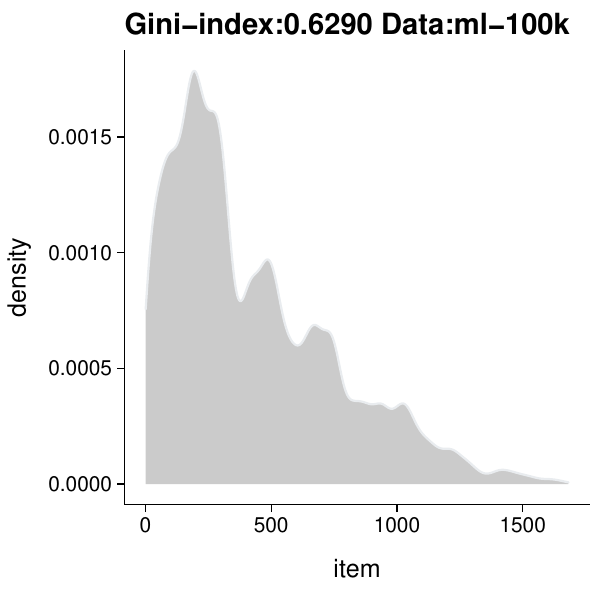}
    \includegraphics[width=4cm]{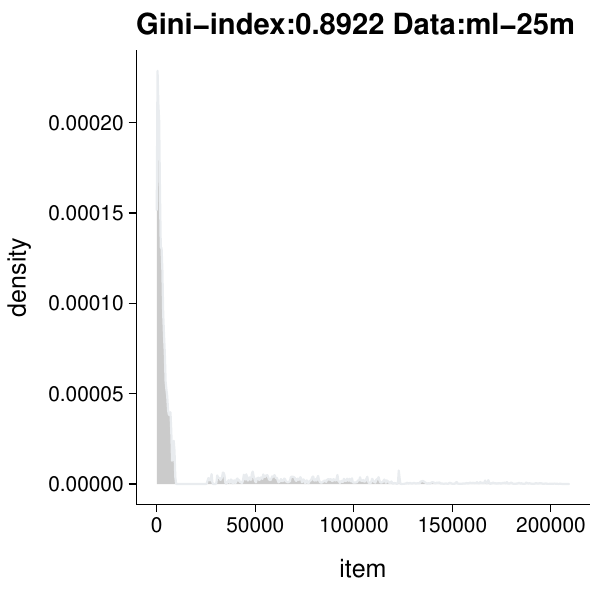}
    \includegraphics[width=4cm]{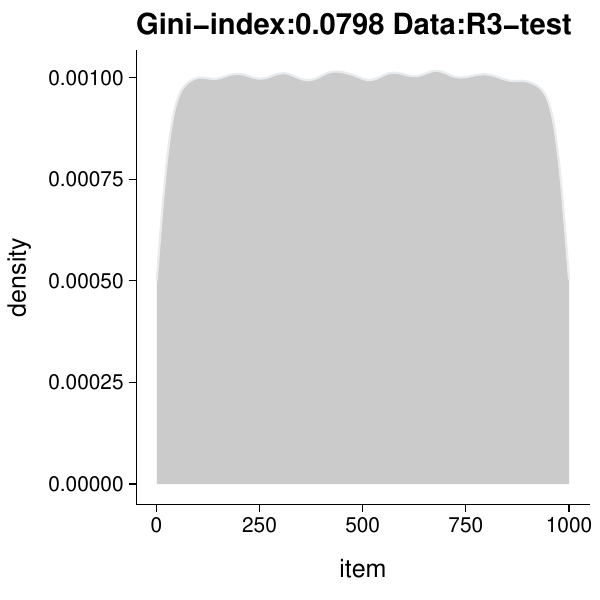}
    \includegraphics[width=4cm]{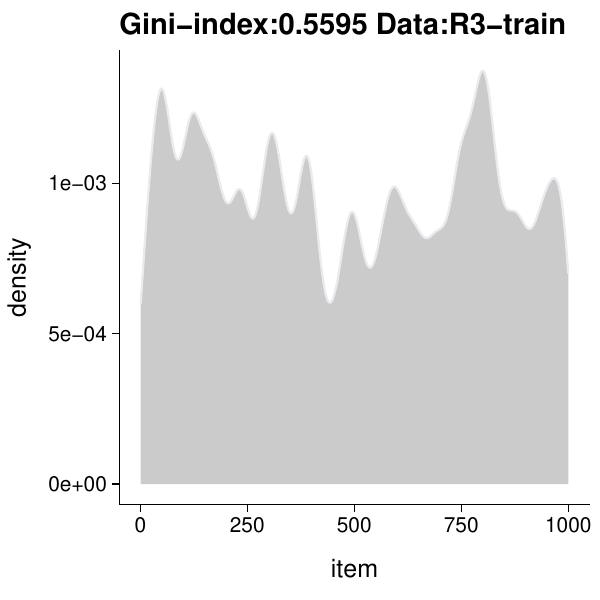}
    \caption{Distribution of ratings over items. The y-axis defines the probability density estimate per unit value of items(x-axis)}
    \label{fig:distr}
\end{figure}

We measured the popularity concentration of ratings in each data set by computing the Gini index of the number of ratings per item, ranging from $0$ (ratings uniformly distributed across items) to $1$ (maximum inequality); see Table \ref{tbl:datasets} and Figure \ref{fig:distr}.

\subsection{Algorithms}
For each of the training partitions, we trained five recommender system algorithms with the default hyper-parameters values using LensKit for Python (version 0.13) \cite{ekstrand2020lenskit}. These include three collaborative filtering algorithms (\textit{Item k-NN (ItemItem)}\cite{deshpande2004item} and \textit{User-based k-NN (UserUser)} \cite{herlocker2002empirical}, both in explicit-feedback mode, along with \textit{Implicit Matrix Factorization ALS (ImplicitMF)}\cite{takacs2011applications}) and two non-personalized algorithms (\textit{Popular} and \textit{Random}). Non-personalized recommenders generate list of items for a user regardless of the user's preferences. They are usually used as baseline for more complex personalized algorithms.

\subsection{Candidate Set Sampling}
We configured and implemented three candidate set selection strategies:
\begin{itemize}
    \item \textbf{full} candidate set: $C_{\User} = \Items \setminus \Items[\User]^{\Ltrain}$ (this is the default for candidate set selection).
    \item \textbf{uniformly-sampled} candidate set: $C_{\User} = \Items[\User]^{\Ltest} \cup \Items[\User]^{\Ldecoy}$ with $\Items[\User]^{\Ldecoy}$ sampled uniformly at random from $\Items \setminus \Items[\User]^{\Ltrain}$.
    \item \textbf{popularity-weighted} candidate set: $C_{\User} = \Items[\User]^{\Ltest} \cup \Items[\User]^{\Ldecoy}$ with each item in $\Items[\User]^{\Ldecoy}$ sampled without replacement from $\Items \setminus \Items[\User]^{\Ltrain}$ with probability proportional to the number of users who have rated it.
\end{itemize}

We composed the candidate sets for different decoy sizes $[10, 20, 50, 100, 200, 500, 1000, 2000]$. We generated ranked recommendations for each test user from each of the candidate set (Full, popularity-weighted, and uniformly sampled).

\subsection{Evaluation Metrics}
\label{eval}
We measured the utility of each recommendation list with $\operatorname{nDCG}$, precision, recall, and reciprocal rank as implemented by LensKit. We used a cutoff of $10$ as the recommendation list metric depth.

We measured the prevalence of popular items in a recommendation list by computing the mean popularity rank: the average of the ranks of recommended items by their popularity (where rank $1$ is the least-popular item). An algorithm's popularity tendency is measured by the average of the mean popularity rank of the lists it produced.

\subsection{Results}

\begin{figure}
    \centering
    \includegraphics[width=\columnwidth]{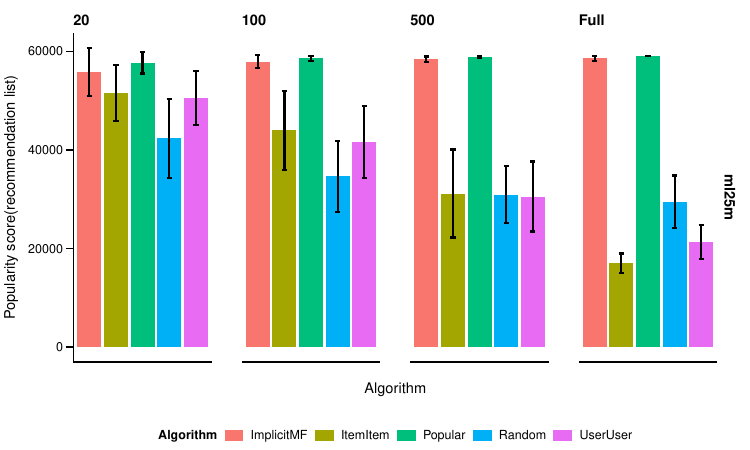}
    \caption{Popularity score is the measure of the prevalence of popular items in the recommendation list produced by each recommender algorithm. The recommendation lists are generated from candidate sets with decoy sizes [20, 100, 500] and the full candidate set.}
    \label{fig:pop-score}
\end{figure}

\begin{figure*}
    \centering
    \includegraphics[width=7cm]{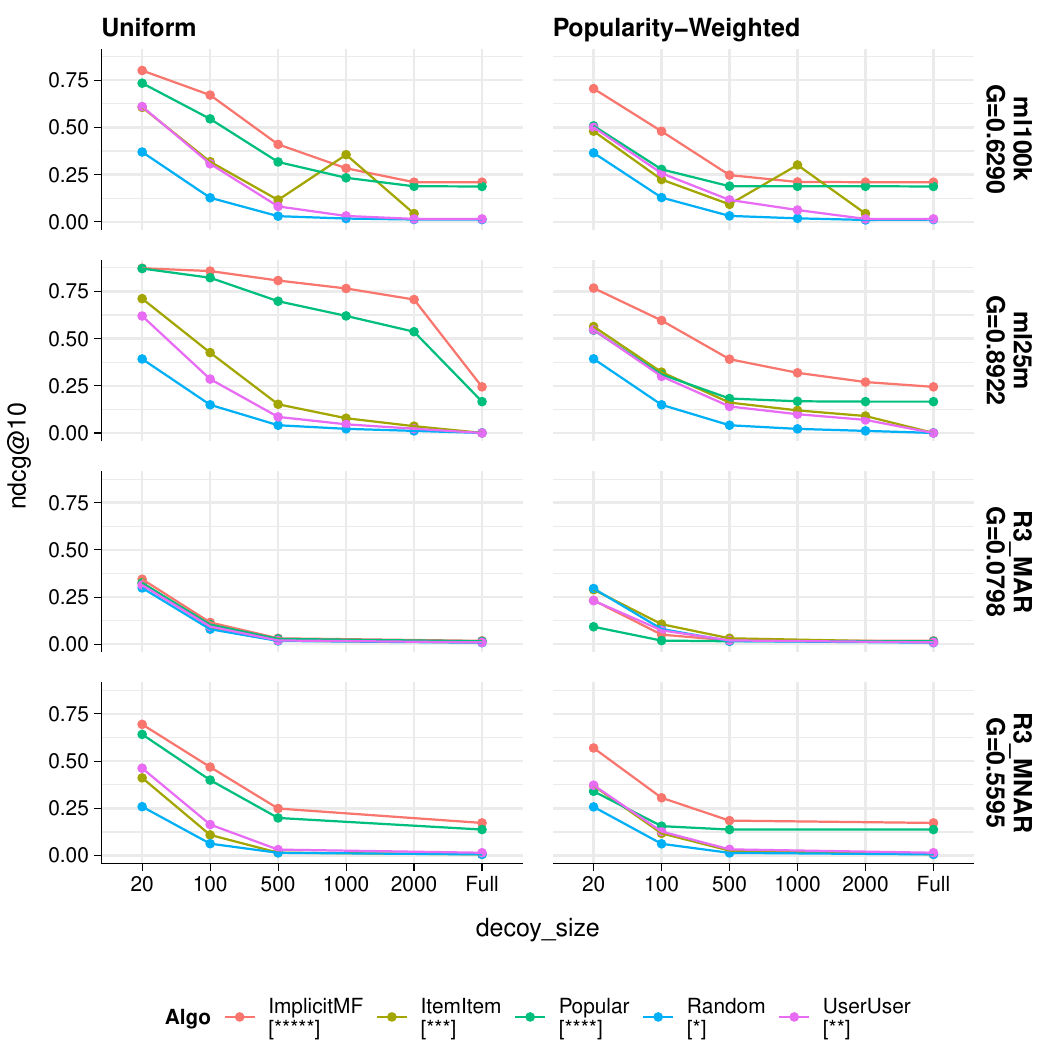}\hspace{1cm}
    \includegraphics[width=7cm]{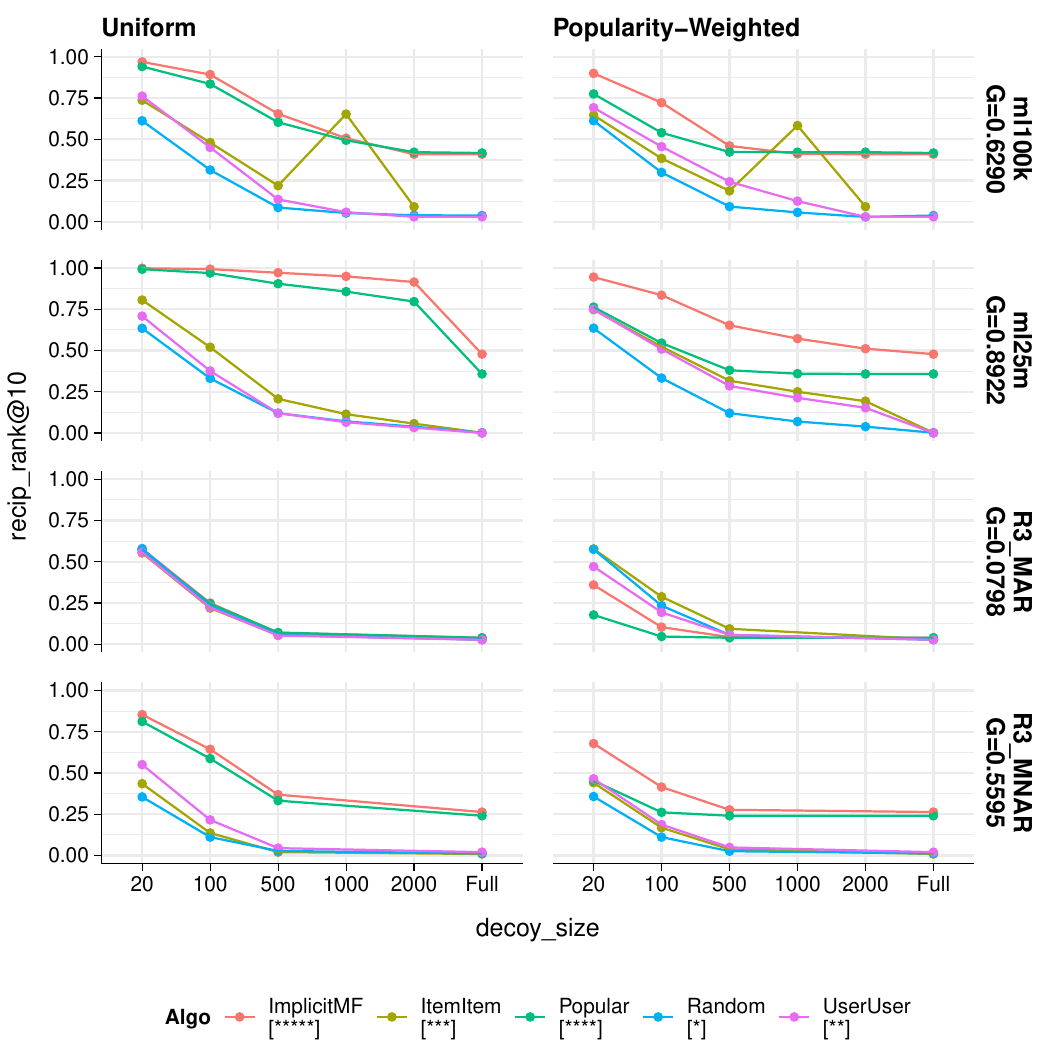}\hspace{1cm}
    \includegraphics[width=7cm]{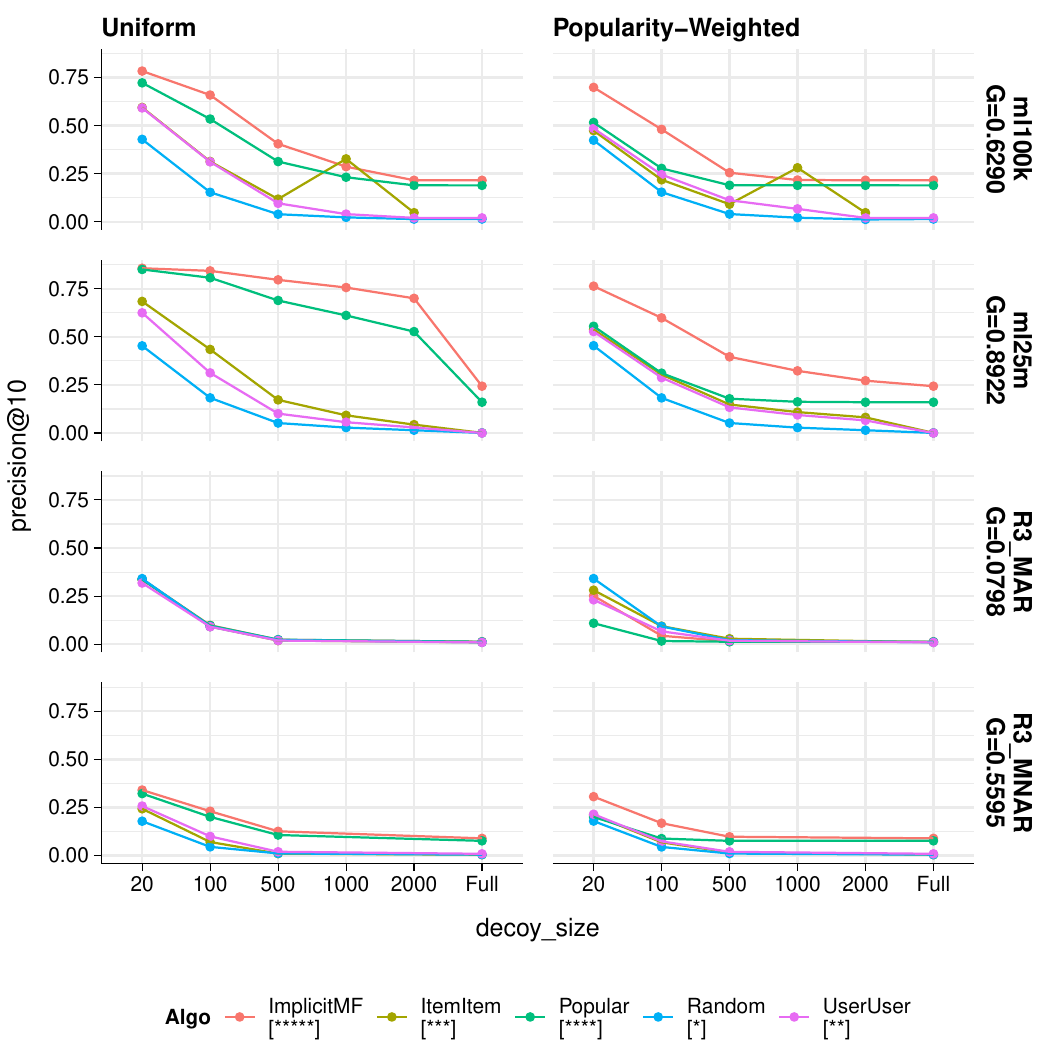}\hspace{1cm}
    \includegraphics[width=7cm]{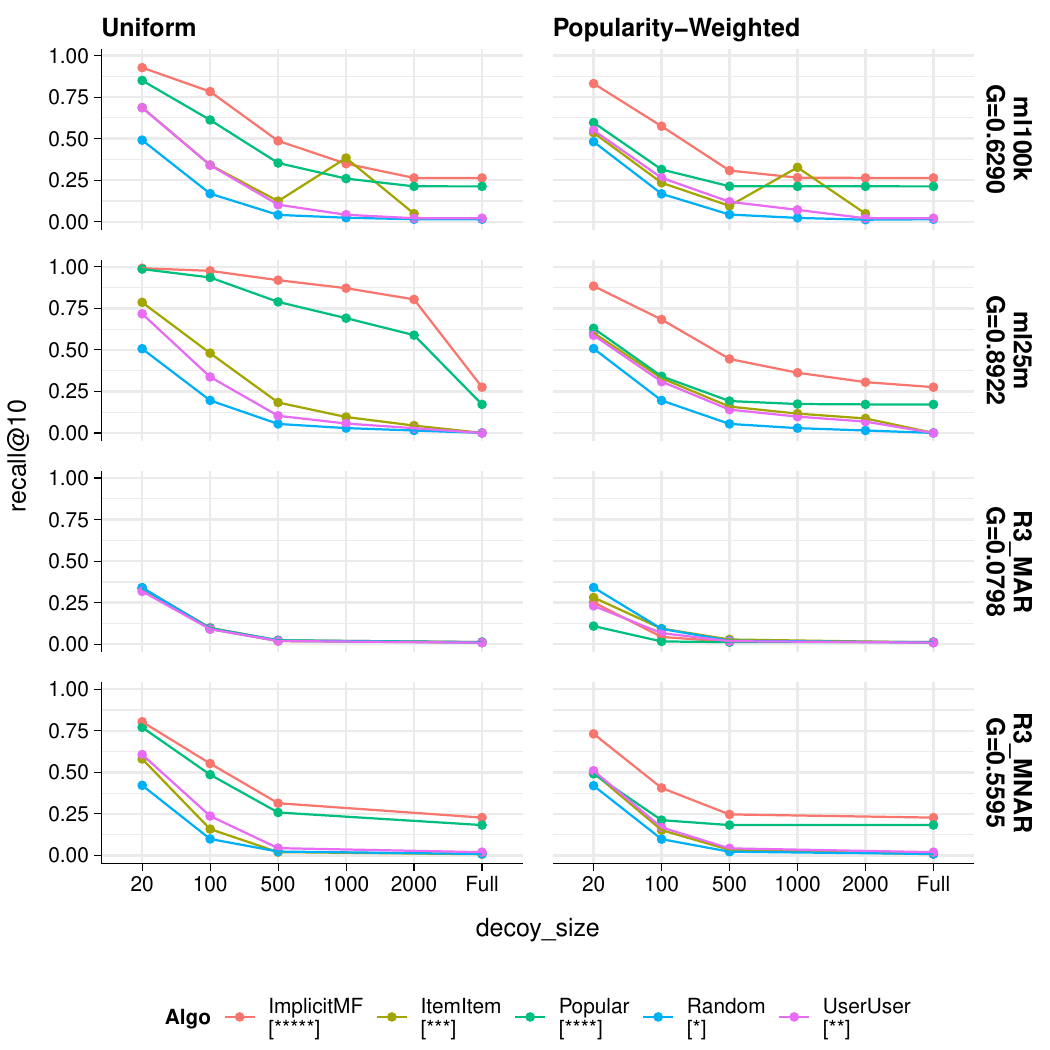}
    \caption{Ranking systems by the $\operatorname{nDCG}$, reciprocal rank, precision and recall metrics computed for recommendation lists produced from popularity-weighted and uniform sampled candidates. The asterisks indicate the relative level of tendency to recommend popular items.}
    \label{fig:sys-rank}
\end{figure*}

In this subsection, we present findings of the combined effect of the factors: candidate set selection strategy, distribution of ratings over items (measured in Gini index), and algorithm popularity tendency on popularity bias. 

First, we define the term \emph{popularity-prone algorithms}. This refers to algorithms that have the tendency to recommend popular items disproportionately to their relevance to users' preferences. This tendency is determined by the prevalence of popular items in the recommendation list they produce as described in subsection \ref{eval}. From Figure \ref{fig:pop-score} we observe very high popularity scores for the ImplicitMF and Popular algorithms across different candidate set sizes. The result shown is based on the ML-25M dataset but we observed similar patterns with other datasets.

Figure \ref{fig:sys-rank} shows the result of the empirical  experiment for $\textbf{RQ1}$ and $\textbf{RQ2}$. As can be seen, the performance of all algorithms declines at approximately similar rates on each of the datasets as the decoy size of the candidate set increases and tends towards the Full set except for those of the popularity-prone algorithms. When the decoys are uniformly sampled, we observe that the performance of popularity-prone algorithms remain consistently high across the decoy sizes, on datasets with high Gini coefficient (ML-25M, Ml-100k, Yahoo!R3[MNAR]). This is especially pronounced on the ML-25M dataset where we observe high performance before we see a sharp drop in performance for the Full set.

We also observe that system ranking remains consistent across datasets. However, the effect sizes decreases as the decoy size increases, and in some cases they are not statistically significant such as on the the Yahoo!R3[MAR] dataset.

\begin{figure*}
    \centering
    \includegraphics[width=7cm]{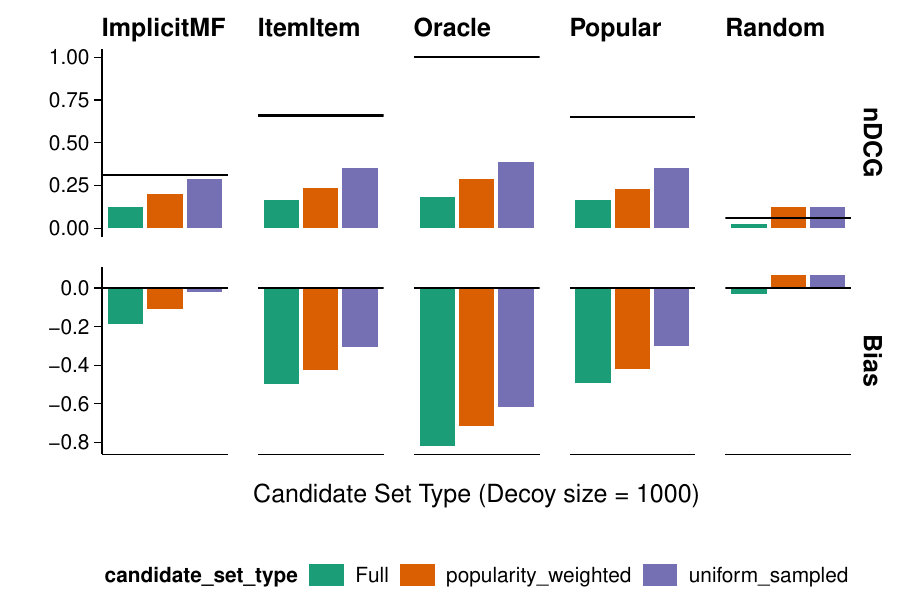}\hspace{1cm}
    \includegraphics[width=7cm]{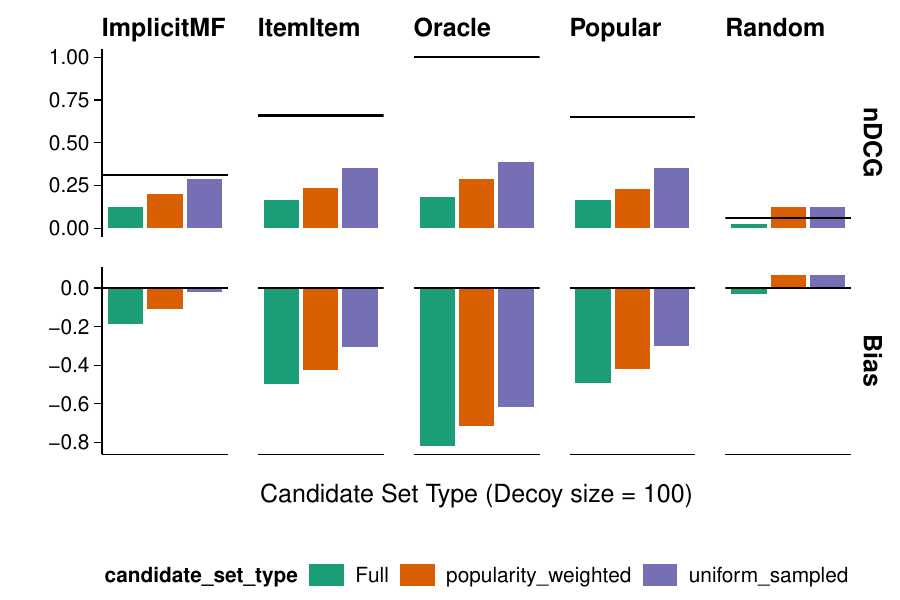}\hspace{1cm}    
    \includegraphics[width=7cm]{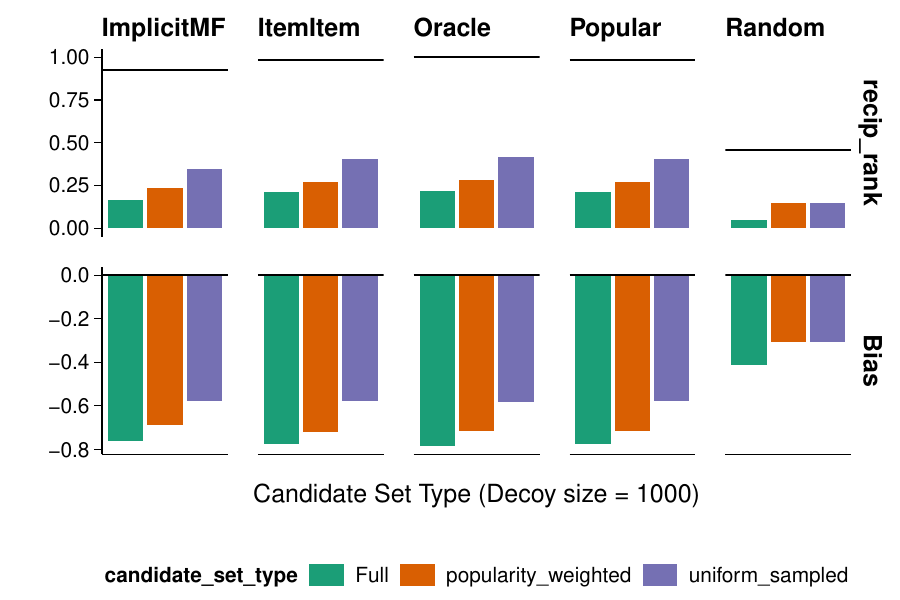}\hspace{1cm}
    \includegraphics[width=7cm]{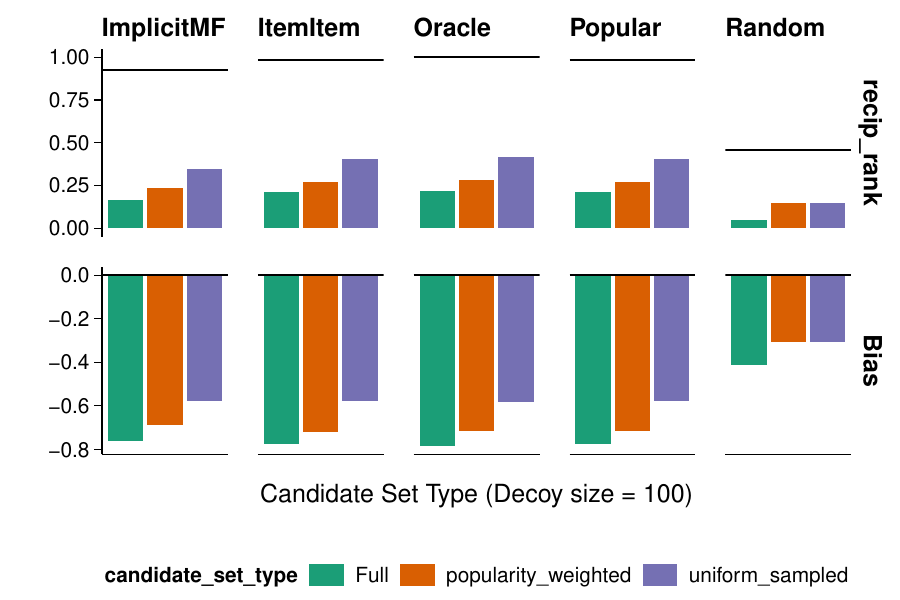}\hspace{1cm}
    \caption{The top row of bar charts show the $\operatorname{nDCG}$ values for recommendation lists generated by each of the algorithms from the full, uniformly sampled, and popularity-weighted candidate sets using the held-out observable test data as ground truth. The horizontal reference lines show the measure ($\operatorname{nDCG}$, reciprocal rank) values for recommendation lists obtained from the full candidate set using the true preference dataset as ground truth.
  The bottom row of bar charts show the bias of the ($\operatorname{nDCG}$, reciprocal rank) metric with reference to the full candidate set. $Bias = M^{obs} - M^{truth}$ }
    \label{fig:sim1}
\end{figure*}
\subsection{Discussion}
In Figure \ref{fig:sys-rank} we see a pattern of behavior by the algorithms that addresses the question: does uniform sampling of the decoys exacerbate popularity bias in evaluation? The performance of the popularity-prone algorithms on the datasets with high Gini coefficient(ML-100K, ML-25M, Yahoo!R3[MNAR]) remain consistently high as the decoy size increases contrarily to the behavior of the other algorithms. This behavior is especially pronounced on the ML-25M dataset which has a Gini coefficient of $0.8922$. A Gini coefficient of $0.8922$ implies that majority of the ratings are concentrated over a few popular items. This implies two things: (1) most items have little or no ratings and are assumed to be non-relevant; (2) majority of the test items are popular because test items are rated. 

We may recall that the candidate set consists of the test items and the decoys. When the decoys are uniformly sampled, it most likely consist of unrated items (assumed to be non-relevant). The evaluation protocol generally assesses the ability of a recommender algorithm to differentiate between the test items and the decoys. The popularity-prone algorithms have a high propensity of selecting popular items; therefore, they are able to pick out the popular items from a collection of unpopular items (unrated or with few ratings) irrespective of how many unpopular items are in the candidate set. Hence, by just picking the popular items they are rewarded by the evaluation protocol. 

We observe a sharp decline in their performance with the Full candidate set. This is because with the Full candidate set it becomes difficult to just pick out the popular items from the unpopular ones. The decoy component of the Full candidate set consists of all items that a user has not interacted with. Hence, the decoys do not mainly consists of unrated unpopular items as is the case with uniformly sampled decoys.

This behavior of the popularity-prone algorithms (ImplicitMF and Popular) on datasets with high Gini index show an exacerbation of the popularity bias in evaluation. This result is in agreement with the analytical findings from \cite{ekstrand2017sturgeon} that the uniform sampling of the decoys may exacerbate popularity bias in evaluation.

We also find patterns in algorithms behavior in Figure \ref{fig:sys-rank} that addresses the question: does popularity-weighted sampling of the decoys mitigate the exacerbation of popularity bias? We observe that when the decoys are sampled by popularity-weighting, all algorithms behave alike: their performance decline as the decoy size increases and tends to the Full candidate set.

Therefore, when the Gini index of a dataset is high, and our algorithm is popularity-prone, we use popularity-weighted sampling of the decoys to ensure that the test and decoy items come from approximately the same marginal distribution. This is demonstrated in figure \ref{fig:sys-rank}, we see that when we sample the decoys by popularity weighting, the popularity prone algorithms are now forced to differentiate between the relevant popular items and the unrated popular items. This behavior is also observed with the Yahoo!R3[MAR] dataset (with Gini index $0.0798$ --- ratings are approximately evenly distributed across items).
Hence, popularity weighting can help mitigate the exacerbation of popularity bias that comes as a result of uniformly sampling the decoys.

\section{Estimating Bias using Simulation}

Having established that popularity-weighting sampled decoys can help reduce the effect of making popularity bias worse, we now turn to $\textbf{RQ3}$, and use a simulation study to assess whether sampled candidate sets --- either uniform or popularity-weighted --- help correct for missing data that induces bias between the metric scores computed with observed data and those that would arise from complete test data if it was available.

Bias, in the statistical sense, is when the estimator differs in expectation from the estimand. In recommender systems metrics, this results in the observed evaluation metric over- or under-estimating the true effectiveness score of an algorithm as it would be computed with a particular metric with complete relevance data available; we are not, in this work, concerned with mapping to online effectiveness measurements. We used simulation to estimate the bias of an evaluation metric with different candidate set selection strategies. The simulation proceeds in three stages: 

\begin{enumerate}
    \item Generate synthetic users and items with preference relations so we know all items a user would like.
    \item Simulate an observation process to observe user preference for a subset of the items they would like (like a real system only sees a subset of a user's preferences).
    \item Run a recommender experiment on the synthetic observations, but have the complete synthetic preference data from (1) available for evaluation. 
\end{enumerate}
 
We adopted the latent Dirichlet allocation ($\operatorname{LDA}$) \cite{blei2003latent} true preference simulation model implemented in  Tian and Ekstrand \cite{tian2020estimating}, and tuned to mimic the ML1M movielens dataset.
$\operatorname{LDA}$ is a Bayesian generative probabilistic model for collections of discrete data such as text corpora. 
It is an example of a latent feature model that provides a mechanism for representing correlations between items. Exploiting these correlations is fundamental to many recommendation techniques. $\operatorname{LDA}$ assumes the following generative process for $K$ latent features \cite[pg. 20 description from original]{tian2019estimating}:

\begin{enumerate}
    \item Draw $K$ feature-item vectors $\phi_K \in [0,1]^{|I|}$ from $\operatorname{Dirichlet}(\beta)$
    \item For each user:
    \begin{enumerate}
        \item Draw a latent feature vector $\theta_u \in [0,1]^{K}$ from $\operatorname{Dirichlet}(\alpha)$
        \item Draw $n_u$ (number of items) from $\operatorname{Poisson}(\lambda)$
        \item Draw items $i_1,..., i_{n_u}$ liked by user $u$ by drawing feature $k_x \sim \operatorname{Multinomial}(\theta_u)$ and $i_x \sim \operatorname{Multinomial}(\phi_{k_x})$
    \end{enumerate}
    \item De-duplicate user-item pairs to produce implicit user preference samples.
\end{enumerate}

\subsection{Experiment}
\label{bias}

\begin{figure}
  \centering
  \includegraphics[width=\columnwidth]{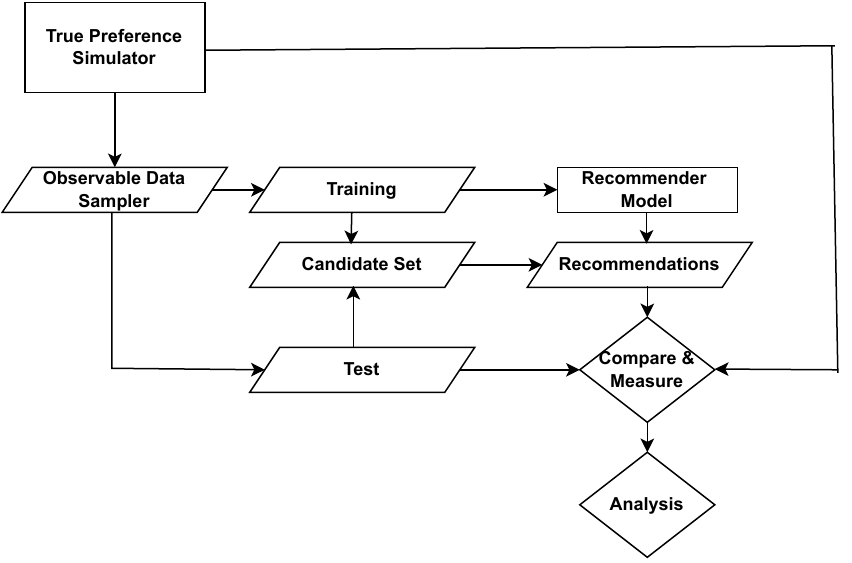}
  \caption{Simulation process for offline experiment}
  \label{fig:sim_proc}
\end{figure}

First, we describe the experiment (see Figure \ref{fig:sim_proc}) that addresses $\textbf{RQ3}$: Does uniform or popularity-weighted sampling of the decoys improve the estimation of effectiveness performance of recommender system in terms of the bias of the computed metric with respect to the value computed over complete relevance data?

To estimate bias, we conduct the following experiment:
\begin{enumerate}
    \item Use the $\operatorname{LDA}$ model to generate the true preference dataset, and from the true preference dataset, we sample sets of observable rated items for each user.
    \item Use the popularity observation sampler implemented in \cite{tian2020estimating} to sample the observations. This sampler operationalizes the idea that users are more likely to rate popular items. 
    \item Split the observations into train and test datasets.
    \item Use the train dataset to train six models: two personalized algorithms (\textit{ImplicitMF} and \textit{ItemItem} in implicit feedback mode); and three non-personalized algorithms (\textit{Popular}, \textit{Oracle}, and \textit{Random}).  The \textit{Oracle} recommender uses the true preference data from (1) to generate perfect recommendations.
    \item Use trained models to generate a $50$-item ranked recommendation list for each test user from each of the candidate sets (full, popularity-weighted, and uniformly sampled), using a decoy size of $1000$.
    \item Compute the $\operatorname{nDCG}$ measure for the recommendation lists generated from the full candidate set by first using the held-out observable test data as ground truth, and again with the simulated true preference data as ground truth thus obtaining the metric values ${M}^\mathrm{obs}_\mathrm{full}$ and $M^\mathrm{truth}$ respectively.
    \item Use the held-out observable test data as ground truth to obtain metric values ${M}^\mathrm{obs}_\mathrm{unif}$ and ${M}^\mathrm{obs}_\mathrm{pop}$ for recommendation lists generated from the uniformly sampled and popularity-weighted candidate sets. 
    \item Compute the bias as the difference between each metric value [${M}^\mathrm{obs}_\mathrm{full}$, ${M}^\mathrm{obs}_\mathrm{unif}$, ${M}^\mathrm{obs}_\mathrm{pop}$] and ${M}^\mathrm{truth}$. 
\end{enumerate}
We repeat the experiment $500$ times.

\subsection{Results and Discussion}

From Figure \ref{fig:sim1} we notice that with the full candidate set, observed data significantly underestimates  the $\operatorname{nDCG}$ and reciprocal rank measures for all algorithms.  The estimation improves with the uniformly sampled, and popularity-weighted candidate sets but is less biased with the uniformly sampled candidate set. We observe that the systems are ranked consistently but the bias varies from algorithm to algorithm with the $\operatorname{nDCG}$ metric. This poses a significant problem for evaluation because if the bias were the same, then we could trust that relative performance is preserved.

\section{conclusion}

The findings of this study provides valuable insight into the effect of candidate set sampling on popularity bias, and their usefulness as a better estimator of effectiveness metric value compared with using the full candidate set. We empirically demonstrated that sampling the decoys of the candidate set uniformly worsens popularity bias in evaluation, and that sampling the decoys by popularity weighting seems to reduce the exacerbation of popularity bias. These findings are consistent with the analytical findings of previous research\cite{ekstrand2017sturgeon} on candidate selection strategies. 

Using simulation, we showed that with the full candidate set, the effectiveness measure for all algorithms is significantly underestimated. However, the estimation improves with the uniformly sampled and popularity-weighted candidate sets indicating the usefulness of candidate set sampling in evaluation. This is in contrast with findings from \cite{krichene2022sampled} that candidate set reduction should not be used because it may affect comparative evaluation. 

Given that popularity weighting seems to reduce the exacerbation of popularity bias, and that estimation improves with sampled candidate set, we recommend that decoys be sampled by popularity-weighting.

However, though the simulation result shows that estimation with the uniformly sampled candidate set is less biased than estimation with popularity-weighted candidate set, this observation needs to be further investigated. We need to ascertain what influence, if any, the high popularity concentration of the ML-1M dataset which the simulation mimics may have had, as well as the impact of limitations in the simulation's ability to mimic the target data set. 

There are many open questions that should be examined in future work, including: effect of other sampling strategies, how big the candidate set should be, the change in bias from system to system, and the bias impact on evaluation.

\bibliographystyle{IEEEtran}
\bibliography{WA-IAT2023}

\begin{thebibliography}{10}
\providecommand{\url}[1]{#1}
\csname url@samestyle\endcsname
\providecommand{\newblock}{\relax}
\providecommand{\bibinfo}[2]{#2}
\providecommand{\BIBentrySTDinterwordspacing}{\spaceskip=0pt\relax}
\providecommand{\BIBentryALTinterwordstretchfactor}{4}
\providecommand{\BIBentryALTinterwordspacing}{\spaceskip=\fontdimen2\font plus
\BIBentryALTinterwordstretchfactor\fontdimen3\font minus
  \fontdimen4\font\relax}
\providecommand{\BIBforeignlanguage}[2]{{%
\expandafter\ifx\csname l@#1\endcsname\relax
\typeout{** WARNING: IEEEtran.bst: No hyphenation pattern has been}%
\typeout{** loaded for the language `#1'. Using the pattern for}%
\typeout{** the default language instead.}%
\else
\language=\csname l@#1\endcsname
\fi
#2}}
\providecommand{\BIBdecl}{\relax}
\BIBdecl

\bibitem{cremonesi2021progress}
P.~Cremonesi and D.~Jannach, ``Progress in recommender systems research:
  Crisis? what crisis?'' \emph{AI Magazine}, vol.~42, no.~3, pp. 43--54, 2021.

\bibitem{canamares2020offline}
R.~Ca{\~n}amares, P.~Castells, and A.~Moffat, ``Offline evaluation options for
  recommender systems,'' \emph{Information Retrieval Journal}, vol.~23, no.~4,
  pp. 387--410, 2020.

\bibitem{ekstrand2017sturgeon}
M.~D. Ekstrand and V.~Mahant, ``Sturgeon and the cool kids: Problems with
  random decoys for top-n recommender evaluation,'' in \emph{The Thirtieth
  International Flairs Conference}, 2017.

\bibitem{koren2008factorization}
Y.~Koren, ``Factorization meets the neighborhood: a multifaceted collaborative
  filtering model,'' in \emph{Proceedings of the 14th ACM SIGKDD international
  conference on Knowledge discovery and data mining}, 2008, pp. 426--434.

\bibitem{zhang2018coupledcf}
Q.~Zhang, L.~Cao, C.~Zhu, Z.~Li, and J.~Sun, ``Coupledcf: Learning explicit and
  implicit user-item couplings in recommendation for deep collaborative
  filtering,'' in \emph{IJCAI International Joint Conference on Artificial
  Intelligence}, 2018.

\bibitem{cheng2021dual}
W.~Cheng, Y.~Shen, L.~Huang, and Y.~Zhu, ``Dual-embedding based deep latent
  factor models for recommendation,'' \emph{ACM Transactions on Knowledge
  Discovery from Data (TKDD)}, vol.~15, no.~5, pp. 1--24, 2021.

\bibitem{ebesu2018collaborative}
T.~Ebesu, B.~Shen, and Y.~Fang, ``Collaborative memory network for
  recommendation systems,'' in \emph{The 41st international ACM SIGIR
  conference on research \& development in information retrieval}, 2018, pp.
  515--524.

\bibitem{bellogin2011precision}
A.~Bellogin, P.~Castells, and I.~Cantador, ``Precision-oriented evaluation of
  recommender systems: an algorithmic comparison,'' in \emph{Proceedings of the
  fifth ACM conference on Recommender systems}, 2011, pp. 333--336.

\bibitem{abdollahpouri2020multi}
H.~Abdollahpouri and M.~Mansoury, ``Multi-sided exposure bias in
  recommendation,'' \emph{arXiv preprint arXiv:2006.15772}, 2020.

\bibitem{krichene2022sampled}
W.~Krichene and S.~Rendle, ``On sampled metrics for item recommendation,''
  \emph{Communications of the ACM}, vol.~65, no.~7, pp. 75--83, 2022.

\bibitem{steck2010training}
H.~Steck, ``Training and testing of recommender systems on data missing not at
  random,'' in \emph{Proceedings of the 16th ACM SIGKDD international
  conference on Knowledge discovery and data mining}, 2010, pp. 713--722.

\bibitem{tian2020estimating}
M.~Tian and M.~D. Ekstrand, ``Estimating error and bias in offline evaluation
  results,'' in \emph{Proceedings of the 2020 Conference on Human Information
  Interaction and Retrieval}, 2020, pp. 392--396.

\bibitem{cremonesi2010performance}
P.~Cremonesi, Y.~Koren, and R.~Turrin, ``Performance of recommender algorithms
  on top-n recommendation tasks,'' in \emph{Proceedings of the fourth ACM
  conference on Recommender systems}, 2010, pp. 39--46.

\bibitem{bellogin2017statistical}
A.~Bellog{\'\i}n, P.~Castells, and I.~Cantador, ``Statistical biases in
  information retrieval metrics for recommender systems,'' \emph{Information
  Retrieval Journal}, vol.~20, pp. 606--634, 2017.

\bibitem{canamares2020target}
R.~Ca{\~n}amares and P.~Castells, ``On target item sampling in offline
  recommender system evaluation,'' in \emph{Proceedings of the 14th ACM
  Conference on Recommender Systems}, 2020, pp. 259--268.

\bibitem{latifi2022sequential}
S.~Latifi, D.~Jannach, and A.~Ferraro, ``Sequential recommendation: A study on
  transformers, nearest neighbors and sampled metrics,'' \emph{Information
  Sciences}, vol. 609, pp. 660--678, 2022.

\bibitem{harper2015movielens}
F.~M. Harper and J.~A. Konstan, ``The movielens datasets: History and
  context,'' \emph{Acm transactions on interactive intelligent systems (tiis)},
  vol.~5, no.~4, pp. 1--19, 2015.

\bibitem{marlin2009collaborative}
B.~M. Marlin and R.~S. Zemel, ``Collaborative prediction and ranking with
  non-random missing data,'' in \emph{Proceedings of the third ACM conference
  on Recommender systems}, 2009, pp. 5--12.

\bibitem{ekstrand2020lenskit}
M.~D. Ekstrand, ``Lenskit for python: Next-generation software for recommender
  systems experiments,'' in \emph{Proceedings of the 29th ACM International
  Conference on Information \& Knowledge Management}, 2020, pp. 2999--3006.

\bibitem{deshpande2004item}
M.~Deshpande and G.~Karypis, ``Item-based top-n recommendation algorithms,''
  \emph{ACM Transactions on Information Systems (TOIS)}, vol.~22, no.~1, pp.
  143--177, 2004.

\bibitem{herlocker2002empirical}
J.~Herlocker, J.~A. Konstan, and J.~Riedl, ``An empirical analysis of design
  choices in neighborhood-based collaborative filtering algorithms,''
  \emph{Information retrieval}, vol.~5, pp. 287--310, 2002.

\bibitem{takacs2011applications}
G.~Tak{\'a}cs, I.~Pil{\'a}szy, and D.~Tikk, ``Applications of the conjugate
  gradient method for implicit feedback collaborative filtering,'' in
  \emph{Proceedings of the fifth ACM conference on Recommender systems}, 2011,
  pp. 297--300.

\bibitem{blei2003latent}
D.~M. Blei, A.~Y. Ng, and M.~I. Jordan, ``Latent dirichlet allocation,''
  \emph{Journal of machine Learning research}, vol.~3, no. Jan, pp. 993--1022,
  2003.

\bibitem{tian2019estimating}
M.~Tian, ``Estimating error and bias of offline recommender system evaluation
  results,'' 2019.

\end{thebibliography}

\end{document}